\documentclass[sigconf, screen]{acmart}

\usepackage[T1]{fontenc}
\usepackage{multirow}
\usepackage{graphicx}
\usepackage{framed}
\usepackage{soul}
\usepackage{textcomp}
\usepackage{float}
\usepackage{todonotes}
\usepackage{import}
\usepackage{color}
\newcommand\rf[1]{\textcolor{black}{#1}}
\usepackage{soul}
\usepackage{pifont}
\usepackage{enumitem}
\usepackage{blindtext}
\usepackage{dirtytalk} 

\newcounter{dfcounter}

\newcounter{odcounter}

\setcopyright{none}

\setcopyright{none}
\renewcommand\footnotetextcopyrightpermission[1]{}
\setcopyright{none}
\makeatletter
\renewcommand\@formatdoi[1]{\ignorespaces}
\makeatother

\begin{document}

\title{Towards a Methodology for Participant Selection in Software Engineering Experiments}
\subtitle{A Vision of the Future}

\author{Valentina Lenarduzzi,$^1$ Oscar Dieste,$^2$ Davide Fucci,$^3$ Sira Vegas$^2$}
\affiliation{
  \institution{$^1$LUT University, Finland  --- $^2$Universidad Politécnica de Madrid, Spain --- $^3$Blekinge Institute of Technology, Sweden } 
}
\email{valentina.lenarduzzi@lut.fi; odieste@fi.upm.es; davide.fucci@bth.se; svegas@fi.upm.es}





\renewcommand{\shortauthors}{Lenarduzzi et al.}

\begin{abstract}
\textbf{\textit{Background.}} Software Engineering (SE) researchers extensively perform experiments with human subjects. Well-defined samples are required to ensure external validity. Samples are selected \textit{purposely} or by \textit{convenience}, limiting the generalizability of results.\\
\textbf{\textit{Objective.}} We aim to depict the current status of participants selection in empirical SE, identifying the main threats and how they are mitigated. We draft a robust approach to participants' selection.\\
\textbf{\textit{Method.}} We reviewed existing participants' selection guidelines in SE, and performed a preliminary literature review to find out how participants' selection is conducted in SE in practice.\\
\textbf{\textit{Results.}} We outline a new selection methodology, by 1) defining the characteristics of the desired population, 2) locating possible sources of sampling available for researchers, and 3) identifying and reducing the ``distance'' between the selected sample and its corresponding population.\\
\textbf{\textit{Conclusion.}} We propose a roadmap to develop and empirically validate the selection methodology.
\end{abstract}



\maketitle




\section{Introduction}
\label{sec:introduction}

Participant selection is an important activity when conducting experiments in Software Engineering (SE)~\cite{wohlin2012experimentation}.
Along with objects selection, it has a direct impact on the generalizability of experimental results. Appropriate participant selection has been traditionally associated with random sampling~\cite[pp. 344-5]{shadish2002experimental}.
If a random sample is representative of the original population, the causal effects found in the sample can be safely generalized to the population~\cite{shadish2002experimental}. 

When the properties of the sample do not resemble the ones of the target population, issues with the representativeness of the participants arise~\cite{baltes2020sampling}, and different validity threats can affect the experimental findings, including \textit{reliability of treatment implementation}, \textit{heterogeneity of subjects}, and \textit{selection}.~\cite{shadish2002experimental}.

However, recruiting the right participants is not easy.
There are \textit{``ethical, political, and logistic constraints often limiting the random selection to less meaningful populations''} \cite[pp. 346]{shadish2002experimental}.
Similarly, \textit{``budget limits the selection of units to a small and geographically circumscribed population at a narrowly prescribed set of places and times''}~\cite[pp. 346]{shadish2002experimental}.
Assembling a cohort\footnote{In statistics, a cohort is a group of subjects who share a defining characteristic or set of characteristics.}
 is a trade-off between having participants working in the same settings and performing the same tasks required by the experiment and having  participants who are \textit{actually} available and willing to participate.


%
%
%

In the simplest instance, a cohort is assembled using the available subjects; it is the case of a ``convenience sample.'' 
Another option is filtering out subjects who do not possess some of the desired characteristic---e.g., industry developers.
Some researchers argue that less experienced subjects, such as students, are valid experiment participants~\cite{salman2015students}.

\rf{It is unclear whether these strategies yield representative samples. Without the representativeness safeguard, the validity of the experimental results cannot be claimed, just argued. Some guidelines have been proposed for surveys \cite{molleri2016survey}, but surveys and experiments have different requirements, and such proposals may not be applicable.}
Participants in SE experiments---as opposed to other fields, such as psychology---require specific skills which are uncommon in the general population (e.g., programming knowledge). These skills are typically hard to measure, and their visible proxies are still uncertain \rf{\cite{deMello2016surveys}} (e.g., are \textit{programming years} equivalent to \textit{programming knowledge?}).

This work is a stepping stone towards improving the current participant selection strategies in SE experiments. 
The end goal of our research agenda is to help experimenters assessing the generalizability of SE experiments results. In particular, we want to answer the following research questions:
\begin{enumerate}
\item [RQ$_1$] Which strategies are currently applied for participant selection in SE experiments?
\item [RQ$_2$] Which population characteristics do researchers pursue and which ones they disregard when sampling?
\item [RQ$_3$] Which factors contribute to the invalid selection of participants? 
\item [RQ$_4$] Which techniques can be applied to improve the selection of SE experiments participants?
\end{enumerate}

This paper is structured as follows: Section~\ref{sec:related-work} presents related work and the state of the art regarding participants selection in experimentation.
Section~\ref{sec:future} shows a possible future, while Section~\ref{sec:Approach} introduces the proposed approach. Our conclusions are presented in Section~\ref{sec:conclusions}.

\section{Background}
\label{sec:related-work}

\subsection{Related Work}


\begin{table*}[t]
\centering
\caption{Threats to validity related to participant selection~\cite{shadish2002experimental}}


\label{tab:threats}
\footnotesize
\begin{tabular}{p{1cm}|p{2.2cm}|p{13.5cm}}
\hline
\textbf{Type} &
  \textbf{Threat} &
  \textbf{Description} \\ \hline
\multirow{3}{*}{\textbf{Statistical}} &
  Reliability of treatment implementation &
  Different people apply the treatment in various ways or the same person on different occasions. \\ \cline{2-3} 
 &
  Heterogeneity of subjects &
  Individual differences cannot be removed entirely. When the sample is too heterogeneous, we may observe such differences rather than the treatment effect. Do not confuse with \textit{selection} below. \\ \hline
\multirow{10}{*}{\textbf{Internal}} &
  Selection &
  The "average" participant differ among treatments, i.e., the experimental groups are not equivalent. This threat typically operates in quasi-experiments and true experiments wrongly randomized or with rather small sample sizes. \\ \cline{2-3} 
 &
  Regression to the mean &
  Subjects that score very high (or very low) on some measure tend to obtain more modest (/better) scores in subsequent evaluations. When the first score has been used to assign subjects to groups (as in some quasi-experiments) or create factor levels (not unusual in true experiments), the treatment effect will be confused with the natural regression towards the mean, i.e., the ''average performance''. \\ \cline{2-3} 
 &
  \textit{Attrition} &
  There are non-random dropouts. For example, different types of participants can drop out of each experimental group, or more may drop out of one group than the other. It is not precisely a participant selection issue, but it may be indirectly related to subjects' characteristics. For instance, students are less likely to drop than other participant types. \\ \hline
\multirow{6}{*}{\textbf{Construct}} &
  \textit{Reactivity to the experimental situation} &
  The subjects react to the experimental situation. This is not directly related to participant selection. For instance, a professional developer may get annoyed by the experimental protocol or uncertain due to the experimental tasks' domain. \\ \cline{2-3} 
\textbf{} &
  \textit{Novelty and disruption effects} &
  Subjects may get excited if the treatments/tasks represent an innovation to them, or behave poorly if they disrupt their routine. Again, not directly related to participant selection, but some profiles, e.g., professionals, may be more heavily affected than others, e.g., students. \\ \hline
\multirow{3}{*}{\textbf{External}} &
  Interaction off the causal relationship with units &
  The effect observed in the experiment is valid for the participating subjects only. This problem arises from sampling from wrong (sub-)populations. For instance, a TDD experiment with volunteer developers attracts TDD fanatics, and thus TDD yields better results. But such results cannot be reproduced in the general developer population. \\ \hline
\end{tabular}%
\end{table*}


Experimental units (human subjects in this manuscript) are never alike. When they are assigned to experimental treatments/conditions, their differences may alter the relationship between the treatments and the response variable in such a way that participants characteristics are confounded with the treatment effect \cite{shadish2002experimental}. 

R.A. Fisher introduced \textit{randomization}---i.e., the stochastic, aleatory assignment of subjects to treatments, as a mechanism to avoid bias and guarantee the validity of the inference tests \cite{hall2007ra}. Fisher worked in an agricultural station, so he probably had no big concerns with the availability of seeds. However, in SE experiments, assembling a cohort of human subjects may be painful due to practical constraints---e.g., the number and type of subjects available\rf{, which might have an influence in the results. Table~\ref{tab:threats} shows the validity threats associated to participant selection~\cite{shadish2002experimental}}.


Generalization has been traditionally associated with random sampling \cite[pp. 342-4]{shadish2002experimental}. Nowadays, the mainstream SE experimental textbooks (e.g., \cite{juristo2013basics,wohlin2012experimentation}) take random sampling as the gold standard for subject selection. A random sample is a sample that has been selected from the population in such a way that every possible item in such sample has an equal probability of being selected \cite[pp. 28]{montgomery2011engineering}. Random sampling in SE will be described in Section~\ref{sec:random_sampling}. However, random sampling is not the most usual sampling approach in SE; convenience sampling is often used. We will discuss convenience sampling on Section
~\ref{ref:convenience}.



\subsubsection{Random sampling in SE}
\label{sec:random_sampling}
Kitchenham \textit{et al.} \cite{kitchenham2002preliminary} provide advice on how to research Empirical (in reality, Experimental) SE. Regarding subject selection, the authors recommend to (1) \textit{identify the population from which the subjects and objects are drawn}, (2) acquire information about the demographic characteristics of the sample, such as their \textit{education background} or \textit{gender}, and (3) \textit{record data about subjects who drop out from the studies}. The purpose of these recommendations is to ensure external validity, perform post-hoc tests (to determine where the statistically differences truly came from) and identify the presence of bias, respectively.

%
%
%
%
%
%

The main issue with Kitchenham \textit{et al.} advice is how to \textit{characterize the population}\rf{, e.g., specifying the skills, experience, or any other desired subject characteristic.}
To the best of our knowledge, there is no set of agreed-upon dimensions that define a population of interest---e.g., software developers. Sampling has utilized proxies for population characteristics which \textit{appear} to be self-evident. 
One example is \textit{programming knowledge}. 
We use proxies such as \textit{occupation}---e.g., student or professional \cite{falessi2018empirical,feldt2018four, shepperd2018inferencing} or \textit{experience years} \cite{vanhanen2005effects,salman2015students}. 
Knowing the distribution of these proxies in the population---e.g., using a survey would make it obvious drawing a representative sample.
However, we do not know with certainty how those proxies relate to programming skills. Some steps in that direction have been taken \cite{Bergersen:2014et}, but they are insufficient to ensure samples' representativeness.

As a matter of fact, \textit{examples of individuals being randomly selected to participate in randomized experiments are rare} \cite[pp. 346]{shadish2002experimental}. A systematic review of controlled experiments by Sj\o berg \textit{et al.} covers 103 primary studies published between 1993 and 2002. The authors report that only two of them claim to have conducted random sampling ~\cite{SHH05} (see Section \ref{sec:statePractice} for further details). Without a well-defined population, it is not possible to make inferences from the sample. Several works have tried to solve this problem. There are two basic solutions: 

\vspace{2mm}

\noindent\textit{2.1.1.a)\hspace{2mm}Acquire the ''right'' demographic characteristics of the sample.}
Sj\o berg \textit{et al.} \cite{sjoberg2002conducting} claim that we need to collect \textit{information about the ability and the variations} among experimental subjects for the results to be generalized. For professionals, they suggest collecting competence, education, experience, age, etc. For students, academic level (undergraduate, postgraduate), subject, age, etc. 

H{\"o}st \textit{et al.} \cite{host2005experimental} propose a classification scheme to decide whether the context factors of different controlled experiments are the same. The factors they propose are the \textit{incentives} and the \textit{experience of subjects}. The latter is clearly a sample characteristic. They suggest the following values:  
\begin{itemize}
\item Undergraduate student with less than three months of recent industrial experience. 
\item Graduate student with less than three months of recent industrial experience.
\item Academic with less than three months of recent industrial experience
\item Any person with recent industrial experience, between 3 months and two years.
\end{itemize}

Nagappan \textit{et al.} \cite{nagappan2013diversity}. introduce a technique called \textit{coverage} to identify software projects of similar characteristics. Coverage depends on a (set of) \textit{similarity functions} which, in turn, are defined on several \textit{dimensions}, such as the \textit{number of developers} or the \textit{total lines of code}. Projects are characterized as points in a $n-$dimension space. Similar projects can be identified using a procedure inspired by clustering. The authors have not applied their approach to experimentation with human subjects. It is highly likely that such dimensions are clearer for software projects than for people since, as we have mention before, the \textit{dimensions} needed to calculate the distance are not available for human subjects yet.



\vspace{2mm}

\noindent\textit{2.1.1.b)\hspace{2mm}Ensure that the pool from which the sample has been drawn is the \say{right} one.}
The process starts with a \textit{source of sampling}, which is a database from which adequate subpopulations of the target population can be systematically retrieved \rf{\cite{deMello2014towards,deMello2016surveys}}. There are several candidates to become a source of sampling: graduation classes, attendees to SE conferences, directories of professional societies, etc.~\rf{\cite{deMello2015characterizing}}. Crowdsourcing platforms, such as Amazon Mechanical Turk and Prolific.ac, are popular among researchers, especially in the social sciences, as ways for recruiting experimental participants~\cite{PS18}.

Next, the \rf{source of sampling} is filtered out to identify the relevant sub-populations. SE research has taken \rf{three} approaches: (1) using a qualification exam, e.g., \cite{stolee2014solving,stolee2016code,stolee2010exploring,stolee2013use,salvaneschi2014empirical}, (2) dropping responses that fall outside of one standard deviation of the mean \cite{fry2012human}\rf{, and (3) creating a search plan which includes a search strategy and inclusion/exclusion criteria \cite{deMello2014towards,deMello2016surveys}.}

Finally, a sampling strategy (simple random sampling, clustering, stratified sampling, systematic sampling, etc.) can be applied to such sources to obtain the experimental subjects.

De Mello \textit{et al.}~\cite{de2015investigatingESEM} compare two samples (from Mechanical Turk and LinkedIn) in an online experiment for evaluating the relevance of Java code snippets to programming tasks\footnote{\rf{A second paper published by De Mello \textit{et al.} \cite{de2015investigatingJSERD} replicate a survey on the characteristics of agility and agile practices in SE \cite{abrantes2013towards}. Being a survey, we do not describe it here. But as a side note, notice that Abrantes \& Travassos \cite{abrantes2013towards} use as source of sampling a list of authors extracted from an existing systematic review \cite{abrantes2007caracterizaccao}. In turn, De Mello \textit{et al.} \cite{de2015investigatingJSERD} use LinkedIn. The results seem to be quite consistent (see also \cite{deMello2014agilidade}), regardless of the source of sampling used.}}. 
The authors find that there are significant differences in the two samples (LinkedIn participants present significantly higher levels of experience in programming and Java than Mechanical Turk), despite a high level of consistency in the experimental results. De Mello \textit{et al.}~\cite{de2015investigatingESEM} concludes that the decision of using MTurk or LinkedIn should be used
\textit{based on the characteristics of the study, the need for more or less experienced participants, expected heterogeneity, and sampling strategy}.

Layman \textit{et al.} \cite{Layman2013using} make several recommendations on using MTurk for SE user studies: qualifications of subjects, data validation, procedure adherence, and independence of observations. Regarding the qualification of subjects, which is our primary interest, they rely on MTurk screening mechanisms---i.e., qualification tests, which they consider satisfactory.

Although MTurk and LinkedIn provide access to a larger sample of participants, these platforms contribute to participants' selection issues. A study in the field of empirical finance~\cite{RRM19} shows how, over time, researchers' concerns about the quality of the data obtained from crowd workers led them to sample participants based on reputation metrics established by the platform. By following this approach, researchers are excluding a large part of the possible sample while repeatedly using the same small group of highly experienced participants for their studies~\cite{PS18}. Likewise, Hauser et al.~\cite{HPC19} present several concerns in using crowdsource workers for research purposes. The use of crowdsourcing platforms exacerbates self-selection bias. On top of deciding to enroll in the platform, participants also decide whether to complete a specific task for the study. According to the authors, this has repercussions on replicability as sampling procedures are challenging to reproduce. 
Finally, Baltes and Diehl~\cite{BD16} also acknowledge that SE findings are based on convenience sampling---for which researchers tend to include participants close to their social and cultural group---and suffer from self-selection bias (e.g., snowball sampling using social media).


\subsubsection{Sampling from convenience populations}\label{ref:convenience}

Given the problems above, it is logical that most SE experiments are conducted using convenience sampling~\cite{sjoberg2002conducting, amir2018there,SHH05} typically (under)graduate students or professionals from companies. A sizable amount of literature discusses the professional-student dichotomy.
A recent paper by Falessi \textit{et al.} \cite{falessi2018empirical} shows some insight into the pros and cons of using students and professionals in experiments. 

According to the authors, different experiments in SE literature show contradictory results as regards the performance of professionals and students. They believe that classifying experiment participants using a binary scale (students/professionals) is not an appropriate approach, and that not considering students as representative of professional developers seems to be an oversimplification, as well as considering the use of professionals in experiments as a panacea. They propose to use subject characterization based on relevant attributes instead. But the characteristics to be used is an open question.

Falessi \textit{et al.} \cite{falessi2018empirical} list several characteristics associated to experiments that use students (vs. using professionals):
\begin{itemize}
    \item Have lower external validity.
    \item Show higher treatment conformance. 
    \item Have no lower relevance, i.e., are not of less interest.
    \item Use convenience sampling (but it is worse in the case of professionals).
    \item Are more willing to learn, and therefore might perform better with new technologies. If professionals are used, the positive effects of a new technology are underestimated.
    \item Have less experience (if experience plays a role, they will perform worse).
    \item Support the improvement of experiment design and protocol better than the use of professionals.
\end{itemize}

The authors propose the following ideas: 

\begin{itemize}

    \item Think about population and validity before experimenting. It allows for a better understanding of the sample being used and the population participants belong to.
    
    \item Consider the population in relation to the research questions.
    
    \item Think about the appropriateness of students participating and the kind of professionals we could compare them with.
    
    \item Once a technology gets promising results in laboratory experiments, it might be time to advance in the path and call for practitioners~\cite{gorschek2006model} 
    
    \item The representativeness of students changes with different contexts. Students are reliable proxies for the developer population of typical start-up environments or SMEs \cite{fagerholm2013platform}, and OSS ecosystems.

    \item Not many highly skilled professionals do the development; most developers become managers after a few years. Therefore, students are reliable proxies.

    \item Papers should provide a detailed description of why the use of students as subjects in the study was appropriate under the specific circumstances. 

\end{itemize}

Finally, they provide a characterization scheme of subject experience based on three dimensions:
\begin{itemize}

	\item Real (to what extent does a subject have real experience?).

	\item Relevant (to what extent is the real experience of a subject relevant?).

	\item Recent (to what extent does a subject have recent relevant experience?).

\end{itemize}

Based on the concepts identified in the previous subsections, we investigated which approaches are adopted in the practice to select participants.



\subsection{State of the Practice}
\label{sec:statePractice}

In order to study the current state of the practice in participants selection, we have first examined how SE experiments select participants using existing secondary studies. Additionally, we have conducted a literature review to screen recent experiments.

\subsubsection{Secondary studies}
We found three studies that characterize and discuss the strategies adopted in the participants selection process~\cite{amir2018there,SHH05,baltes2020sampling}. 

Amir and Ralph~\cite{amir2018there} investigated 236 papers published at EMSE between 2012 and 2016.
They show that actual random sampling is rare (only 3\% of the studies), whereas purposive (i.e., create a sample based on a strategy) and convenience (i.e., create a sample arbitrarily) sampling are dominant and stable over time. 
The authors attribute the lack of random sampling to the lack of well-defined sampling frames from which elements can be selected.\footnote{Another assumption is that a random sampling strategy will result in a representative bias. For an in depth explanation of why this reasoning is flawed, we refer to \cite{baltes2020sampling, amir2018there}.}

Purposive sampling is the dominant strategy in SE experiments also according to a review of the literature by Baltes and Raph~\cite{baltes2020sampling} which includes 115 empirical studies published between 2014 and 2019 at the International Conference on Software Engineering (ICSE),  ACM Joint European Software Engineering Conference and Symposium on the Foundations of Software Engineering (ESEC/FSE), Transaction in Software Engineering (TSE), and Transaction in Software Engineering and Methodology (TOSEM). 
The authors show that purposive sampling and random sampling are respectively used in 77\% and 4\% of the experiments.
Other sampling techniques, stratified sampling is used in 1\% of the experiments whereas snowballing is not used at all.

Sj\o berg \textit{et al.} review shows that only two (out of 103) primary studies claim random sampling, but how it was carried out was not reported~\cite{SHH05}. 
The majority of the experiments use convenience sampling, with students being the most used type of participants~\cite{SHH05}.
The authors recommend to incentivize professionals to be part of experimental studies, for example, by compensating employers for the hours their personnel spent on a study, as well as offer workshops to the companies where assessments are used as experimental session~\cite{SHH05,vegas2015difficulties}.

\subsubsection{Literature review}
In the remaining part of this section, we report the selection process we adopted to collect data from the existing literature. 


\textit{Search and Selection Process and Data Extraction.}
As recommended by Sj\'{o}berg~\cite{SHH05}, we used the search string: 
\begin{center}
``\textit{controlled experiment*}'' OR ``\textit{controlled-experiment*}'' OR ``\textit{randomize control trial}''  OR ``\textit{group comparison}''
\end{center}

\rf{We used the asterisk character (*) to capture possible term variations such as plurals.} We applied this string on title and abstract limiting the search to the SE domain. 

\rf{We selected the list of relevant bibliographic sources following Kitchenham and Charters~\cite{Kitchenham2007} suggestions, since these sources are recognized as the most representative in the SE domain and used in many reviews\footnote{The list includes: \textit{ACM Digital Library, IEEEXplore Digital Library, Science Direct, Scopus, Google Scholar, CiteSeer library, Inspec, Springer link}}.  Moreover, we performed a manual search on the most important conferences and journals on Empirical SE, such as  Empirical Software Engineering Journal and International Symposium on Empirical Software Engineering and Measurement (ESEM).}

\rf{We defined inclusion and exclusion criteria to be applied to the title and abstract or to the full text. The retrieved papers should be empirical studies~\cite{wohlin2014guidelines}, had to clearly report the methodology and in particular the case and subject selection.} 

\rf{The search was conducted in March/April 2021 and included all the publications available until this period.} Moreover, we decided to concentrate only on the papers published in the last five years (2017-2021), retrieving 118 unique papers. \footnote{The raw data are available here: https://figshare.com/s/efcd14e2c74d96670515}.
\rf{We tested the applicability of inclusion and exclusion criteria on a subset of ten papers (assigned to all the authors) randomly selected from the papers retrieved.} \rf{Moreover, we performed the snowballing process, applying the same process as for the retrieved papers.}

For each paper, we checked two aspects: 1) the strategy adopted for participants selection and 2) the identified validity threats related to this selection.

\textit{Results}
\textit{Participants selection strategy:}
The vast majority of the considered studies (90 studies) selected the participants based on the current availability at the moment of the study. While the remaining studies did not report any details about the selection. 
In case of bachelor or master students, the authors selected them from the closer related course to the topic of the study. 
In other cases, the authors selected the participants based on their experience in the study context, such as programming languages.
When participants characteristics were taken into consideration, the authors simply tested their experience in the domain profiling the background skills via preliminary surveys for industrial participants.





\textit{Threats to validity for participant selection:} In only 50 papers out of the 118 retrieved ones, the authors \rf{mentioned possible limitations due to participants' selection. In particular, the unanimously reported issue was that selecting other participants with different skills or expertise might have resulted in different outcomes.}  However, none of them discussed possible mitigation actions.

\section{A Possible Future}
\label{sec:future}

\rf{The state of practice presented in Section~\ref{sec:related-work} provides evidence regarding two main issues with participants selection in SE experiments:} 1) we do not know which are the relevant characteristics to define a sample/population properly, and 2) existing source like Mechanical Turk or LinkedIn are less homogeneous than they seem. As a consequence, the most used participants selection methods are \textit{purposive} and \textit{convenience} sampling. These two methods have issues that affect experiment validity.

The goal of our research is to \textbf{propose a methodology that allows researchers to mitigate these validity issues}.





A starting point for the selection methodology is the set of possible scenarios for SE experimentation. \rf{Based on the strategies adopted and threats identified in the literature Section~\ref{sec:statePractice}, as well as on our experience running experiments (both in academia and industry) for more than two decades}, three scenarios are apparent:
\begin{enumerate}
\item Experiments may have \textit{universal aims and goals}---e.g., comparing two refactoring tools for the Java programming language. The \textit{population is very loosely defined}---e.g., subjects with Java programming knowledge.
\item SE experimentation is frequently opportunistic. Industry experiments are a typical example. The \textit{population is given} (the company employees) and quite often the \textit{sample is also pre-defined}---e.g., volunteers. The experiment topic also needs to align with the preferences of the company. 
\item Experiments may have \textit{specific goals} that require a very \textit{well-defined population}---e.g., comparing two control-flow test case design techniques from the testers' viewpoint.
\end{enumerate}

These scenarios suggest the requirements for any participants' selection methodology:
\begin{itemize}
\item We shall be able to estimate to what extent the sample \say{covers} the target population\footnote{This requirement represents the ideal case that is satisfied with random sampling. But SE experiments frequently belong fall in scenarios 2-3.} (scenario 1).
\item We cannot miss the opportunity of running experiments because we do not have the \say{ideal} sample (scenario 2).
\item We shall be able to evaluate the generalizability of the experimental results on the grounds of the data collected from the sample (scenarios 2-3).
\item If the sample departs \say{strongly} from the \say{ideal} population,  we shall be able to evaluate whether running the experiment is worth (scenarios 2-3).
\end{itemize}

Two mechanisms could fulfill these requirements: 1) a procedure to characterize the population/sample (in line with Bergersen \textit{et al.} \cite{Bergersen:2014et} research) and 2) a procedure to evaluate the similarity among samples, or samples vs. populations (similar to the approach proposed by Nagappan \textit{et al.} \cite{nagappan2013diversity} for mining software repository studies). Accordingly, our proposal suggests the following steps:

\begin{itemize}
    
    \item  Identify the characteristics of the desired population and the samples that could potentially be obtained from it.
    
    \item Locate possible sources of sampling  available for the researchers. 
    
    \item Identify the \say{distance} between sample and population, and/or among samples (in terms of their characteristics).
    
    \item Identify how the distance can be reduced, and try to reduce it as much as possible. 
    
    \item Assess to what extent the results support the initial assumptions about the sample and interpret the results accordingly.
    
\end{itemize}
\section{Towards a New Methodology}
\label{sec:Approach}



Our proposed roadmap includes the following steps: 

\noindent\textbf{Step 1: Identification of current approaches, main issues, and mitigation strategies}. We will conduct a systematic literature review to investigate the main approaches adopted by the SE community for participants selection, related issues, and approaches adopted to mitigate them.
 
\noindent\textbf{Step 2: New methodology proposal}. Based on the results of the previous step, we will define a participant selection methodology that will address the aforementioned issues. As mentioned in Section~\ref{sec:future}, it will be based on identifying the \textit{distance} between the current sample and the target population, and apply strategies to reduce this distance to the minimum. However, we need to address the following issues:

\begin{itemize}
    \item How to identify the relevant characteristics of the sample/population.
    \item How to select the most appropriate instruments to measure these characteristics.
    \item How to measure the distance between the sample and the population.
    \item What are the strategies that could be used to reduce the distance.
\end{itemize}

When addressing the identified issues, we may consider some recommendations proposed by \cite{Shadish2001} for causal generalization: 

\begin{itemize}
    \item \textbf{Surface similarity.} In order to guarantee the experimental subjects matching with the important prototypical characteristics of the target population.
    \item \textbf{Ruling out irrelevancies.} In order to check if the participant variation is irrelevant to the size/direction of the effect
    \item \textbf{Interpolation and extrapolation.} In order to check the low and high end of the people's prototypical characteristics.
    \item \textbf{Causal explanation} of the achieved results.
\end{itemize}



\noindent\textbf{Step 3: Internal Validation.} We aim at validating our approach by replicating previous studies. We will select studies from Step 1 which the authors of this paper co-authored.
We will compare the accuracy of our methodology with the current approaches via controlled experiments. 

\noindent\textbf{Step 4: External Validation}. We will involve some authors of the studies considered in Step 1 in order to collect their feedback. Moreover, we will ask them to adopt our methodology in some of their future work.
We will not explain any details in order to avoid bias influencing the researchers.


\section{Conclusions} \label{sec:conclusions}
Despite Empirical SE relying heavily on experimentation, issues  with participants selection are impeding, among others, the correct application of experiments findings in practice---e.g., in terms of generalizability from the experimental sample to the population of interest.
In this paper, we present a detailed overview of the issues related to participants selection, and their impact on several threats to experiment validity. 
Moreover, we present the state-of-practice of participants selection strategies in SE experiments. 
Taking the previous into account, we propose i) a new methodology for participants selection that mitigates validity issues, and ii) a roadmap for its development and evaluation.

\bibliographystyle{ACM-Reference-Format}

\balance

\bibliography{Manuscript.bib}


\begin{thebibliography}{43}


\ifx \showCODEN    \undefined \def \showCODEN     #1{\unskip}     \fi
\ifx \showDOI      \undefined \def \showDOI       #1{#1}\fi
\ifx \showISBNx    \undefined \def \showISBNx     #1{\unskip}     \fi
\ifx \showISBNxiii \undefined \def \showISBNxiii  #1{\unskip}     \fi
\ifx \showISSN     \undefined \def \showISSN      #1{\unskip}     \fi
\ifx \showLCCN     \undefined \def \showLCCN      #1{\unskip}     \fi
\ifx \shownote     \undefined \def \shownote      #1{#1}          \fi
\ifx \showarticletitle \undefined \def \showarticletitle #1{#1}   \fi
\ifx \showURL      \undefined \def \showURL       {\relax}        \fi
\providecommand\bibfield[2]{#2}
\providecommand\bibinfo[2]{#2}
\providecommand\natexlab[1]{#1}
\providecommand\showeprint[2][]{arXiv:#2}

\bibitem[\protect\citeauthoryear{Abrantes and Travassos}{Abrantes and
  Travassos}{2007}]%
        {abrantes2007caracterizaccao}
\bibfield{author}{\bibinfo{person}{Jos{\'e}~Fortuna Abrantes} {and}
  \bibinfo{person}{Guilherme~Horta Travassos}.}
  \bibinfo{year}{2007}\natexlab{}.
\newblock \showarticletitle{Caracteriza{\c{c}}{\~a}o de m{\'e}todos {\'A}geis
  de desenvolvimento de software}. In \bibinfo{booktitle}{\emph{Simp{\'o}sio
  Brasileiro de Qualidade de Software}}.
\newblock


\bibitem[\protect\citeauthoryear{Abrantes and Travassos}{Abrantes and
  Travassos}{2013}]%
        {abrantes2013towards}
\bibfield{author}{\bibinfo{person}{Jos{\'e}~Fortuna Abrantes} {and}
  \bibinfo{person}{Guilherme~Horta Travassos}.}
  \bibinfo{year}{2013}\natexlab{}.
\newblock \showarticletitle{Towards pertinent characteristics of agility and
  agile practices for software processes}.
\newblock \bibinfo{journal}{\emph{CLEI Electronic Journal}}
  \bibinfo{volume}{16}, \bibinfo{number}{1} (\bibinfo{year}{2013}),
  \bibinfo{pages}{6--6}.
\newblock


\bibitem[\protect\citeauthoryear{Amir and Ralph}{Amir and Ralph}{2018}]%
        {amir2018there}
\bibfield{author}{\bibinfo{person}{Bilal Amir} {and} \bibinfo{person}{Paul
  Ralph}.} \bibinfo{year}{2018}\natexlab{}.
\newblock \showarticletitle{There is no random sampling in software engineering
  research}. In \bibinfo{booktitle}{\emph{Int. Conf. on Software Engineering:
  Companion}}. \bibinfo{pages}{344--345}.
\newblock


\bibitem[\protect\citeauthoryear{Baltes and Diehl}{Baltes and Diehl}{2016}]%
        {BD16}
\bibfield{author}{\bibinfo{person}{Sebastian Baltes} {and}
  \bibinfo{person}{Stephal Diehl}.} \bibinfo{year}{2016}\natexlab{}.
\newblock \showarticletitle{{Worse Than Spam}}.
\newblock \bibinfo{journal}{\emph{Int. Symp. on Empirical Software Engineering
  and Measurement}} (\bibinfo{year}{2016}), \bibinfo{pages}{1--6}.
\newblock


\bibitem[\protect\citeauthoryear{Baltes and Ralph}{Baltes and Ralph}{2020}]%
        {baltes2020sampling}
\bibfield{author}{\bibinfo{person}{Sebastian Baltes} {and}
  \bibinfo{person}{Paul Ralph}.} \bibinfo{year}{2020}\natexlab{}.
\newblock \showarticletitle{Sampling in software engineering researcansen: A
  critical review and guidelines}.
\newblock \bibinfo{journal}{\emph{arXiv preprint arXiv:2002.07764}}
  (\bibinfo{year}{2020}).
\newblock


\bibitem[\protect\citeauthoryear{Bergersen, Sjoberg, and Dyba}{Bergersen
  et~al\mbox{.}}{2014}]%
        {Bergersen:2014et}
\bibfield{author}{\bibinfo{person}{Gunnar~R. Bergersen}, \bibinfo{person}{Dag
  I.~K. Sjoberg}, {and} \bibinfo{person}{Tore Dyba}.}
  \bibinfo{year}{2014}\natexlab{}.
\newblock \showarticletitle{{Construction and Validation of an Instrument for
  Measuring Programming Skill}}.
\newblock \bibinfo{journal}{\emph{IEEE Transactions on Software Engineering}}
  \bibinfo{volume}{40}, \bibinfo{number}{12} (\bibinfo{year}{2014}),
  \bibinfo{pages}{1163--1184}.
\newblock


\bibitem[\protect\citeauthoryear{de~Mello, da~Silva, Runeson, and
  Travassos}{de~Mello et~al\mbox{.}}{2014b}]%
        {deMello2014towards}
\bibfield{author}{\bibinfo{person}{Rafael~Maiani de Mello},
  \bibinfo{person}{Pedro~Correa da Silva}, \bibinfo{person}{Per Runeson}, {and}
  \bibinfo{person}{Guilherme~Horta Travassos}.}
  \bibinfo{year}{2014}\natexlab{b}.
\newblock \showarticletitle{Towards a Framework to Support Large Scale Sampling
  in Software Engineering Surveys}. In \bibinfo{booktitle}{\emph{Int. Symp. on
  Empirical Software Engineering and Measurement}}.
\newblock


\bibitem[\protect\citeauthoryear{de~Mello, da~Silva, and Travassos}{de~Mello
  et~al\mbox{.}}{2014a}]%
        {deMello2014agilidade}
\bibfield{author}{\bibinfo{person}{Rafael~M de Mello}, \bibinfo{person}{Pedro~C
  da Silva}, {and} \bibinfo{person}{Guilherme~H Travassos}.}
  \bibinfo{year}{2014}\natexlab{a}.
\newblock \showarticletitle{Agilidade em Processos de Software: Evid{\^e}ncias
  Sobre Caracter{\'\i}sticas de Agilidade e Pr{\'a}ticas {\'A}geis}. In
  \bibinfo{booktitle}{\emph{Simp{\'o}sio Brasileiro de Qualidade de Software}}.
  SBC, \bibinfo{pages}{151--165}.
\newblock


\bibitem[\protect\citeauthoryear{de~Mello, Da~Silva, and Travassos}{de~Mello
  et~al\mbox{.}}{2015a}]%
        {de2015investigatingJSERD}
\bibfield{author}{\bibinfo{person}{Rafael~Maiani de Mello},
  \bibinfo{person}{Pedro~Corr{\^e}a Da~Silva}, {and}
  \bibinfo{person}{Guilherme~Horta Travassos}.}
  \bibinfo{year}{2015}\natexlab{a}.
\newblock \showarticletitle{Investigating probabilistic sampling approaches for
  large-scale surveys in software engineering}.
\newblock \bibinfo{journal}{\emph{Journal of Software Engineering Research and
  Development}} \bibinfo{volume}{3}, \bibinfo{number}{1}
  (\bibinfo{year}{2015}), \bibinfo{pages}{1--26}.
\newblock


\bibitem[\protect\citeauthoryear{de~Mello, Stolee, and Travassos}{de~Mello
  et~al\mbox{.}}{2015b}]%
        {de2015investigatingESEM}
\bibfield{author}{\bibinfo{person}{Rafael~M de Mello},
  \bibinfo{person}{Kathryn~T Stolee}, {and} \bibinfo{person}{Guilherme~H
  Travassos}.} \bibinfo{year}{2015}\natexlab{b}.
\newblock \showarticletitle{Investigating samples representativeness for an
  online experiment in java code search}. In \bibinfo{booktitle}{\emph{Int.
  Symp. on Empirical Software Engineering and Measurement}}.
  \bibinfo{pages}{1--10}.
\newblock


\bibitem[\protect\citeauthoryear{de~Mello and Travassos}{de~Mello and
  Travassos}{2015}]%
        {deMello2015characterizing}
\bibfield{author}{\bibinfo{person}{Rafael~Maiani de Mello} {and}
  \bibinfo{person}{Guilherme~Horta Travassos}.}
  \bibinfo{year}{2015}\natexlab{}.
\newblock \showarticletitle{Characterizing Sampling Frames in Software
  Engineering Surveys.}. In \bibinfo{booktitle}{\emph{CIbSE}}.
  \bibinfo{pages}{267}.
\newblock


\bibitem[\protect\citeauthoryear{de~Mello and Travassos}{de~Mello and
  Travassos}{2016}]%
        {deMello2016surveys}
\bibfield{author}{\bibinfo{person}{Rafael~Maiani de Mello} {and}
  \bibinfo{person}{Guilherme~Horta Travassos}.}
  \bibinfo{year}{2016}\natexlab{}.
\newblock \showarticletitle{Surveys in Software Engineering: Identifying
  Representative Samples}. In \bibinfo{booktitle}{\emph{Int.Symp. on Empirical
  Software Engineering and Measurement}}. Article \bibinfo{articleno}{55},
  \bibinfo{numpages}{6}~pages.
\newblock


\bibitem[\protect\citeauthoryear{Fagerholm, Oza, and M{\"u}nch}{Fagerholm
  et~al\mbox{.}}{2013}]%
        {fagerholm2013platform}
\bibfield{author}{\bibinfo{person}{Fabian Fagerholm}, \bibinfo{person}{Nilay
  Oza}, {and} \bibinfo{person}{J{\"u}rgen M{\"u}nch}.}
  \bibinfo{year}{2013}\natexlab{}.
\newblock \showarticletitle{A platform for teaching applied distributed
  software development: The ongoing journey of the Helsinki software factory}.
  In \bibinfo{booktitle}{\emph{Int. Workshop on Collaborative Teaching of
  Globally Distributed Software Development}}. \bibinfo{pages}{1--5}.
\newblock


\bibitem[\protect\citeauthoryear{Falessi and et~al.}{Falessi and
  et~al.}{2018}]%
        {falessi2018empirical}
\bibfield{author}{\bibinfo{person}{Davide Falessi} {and} \bibinfo{person}{et
  al.}} \bibinfo{year}{2018}\natexlab{}.
\newblock \showarticletitle{Empirical software engineering experts on the use
  of students and professionals in experiments}.
\newblock \bibinfo{journal}{\emph{Empirical Software Engineering}}
  \bibinfo{volume}{23}, \bibinfo{number}{1} (\bibinfo{year}{2018}),
  \bibinfo{pages}{452--489}.
\newblock


\bibitem[\protect\citeauthoryear{Feldt and et~al.}{Feldt and et~al.}{2018}]%
        {feldt2018four}
\bibfield{author}{\bibinfo{person}{Robert Feldt} {and} \bibinfo{person}{et
  al.}} \bibinfo{year}{2018}\natexlab{}.
\newblock \showarticletitle{Four commentaries on the use of students and
  professionals in empirical software engineering experiments}.
\newblock \bibinfo{journal}{\emph{Empirical Software Engineering}}
  \bibinfo{volume}{23}, \bibinfo{number}{6} (\bibinfo{year}{2018}),
  \bibinfo{pages}{3801--3820}.
\newblock


\bibitem[\protect\citeauthoryear{Fry, Landau, and Weimer}{Fry
  et~al\mbox{.}}{2012}]%
        {fry2012human}
\bibfield{author}{\bibinfo{person}{Zachary~P Fry}, \bibinfo{person}{Bryan
  Landau}, {and} \bibinfo{person}{Westley Weimer}.}
  \bibinfo{year}{2012}\natexlab{}.
\newblock \showarticletitle{A human study of patch maintainability}. In
  \bibinfo{booktitle}{\emph{Int. Symp. on Software Testing and Analysis}}.
  \bibinfo{pages}{177--187}.
\newblock


\bibitem[\protect\citeauthoryear{Gorschek, Garre, Larsson, and Wohlin}{Gorschek
  et~al\mbox{.}}{2006}]%
        {gorschek2006model}
\bibfield{author}{\bibinfo{person}{Tony Gorschek}, \bibinfo{person}{Per Garre},
  \bibinfo{person}{Stig Larsson}, {and} \bibinfo{person}{Claes Wohlin}.}
  \bibinfo{year}{2006}\natexlab{}.
\newblock \showarticletitle{A Model for Technology Transfer in Practice}.
\newblock \bibinfo{journal}{\emph{IEEE Software}} \bibinfo{volume}{23},
  \bibinfo{number}{6} (\bibinfo{year}{2006}), \bibinfo{pages}{88--95}.
\newblock


\bibitem[\protect\citeauthoryear{Hall}{Hall}{2007}]%
        {hall2007ra}
\bibfield{author}{\bibinfo{person}{Nancy~S Hall}.}
  \bibinfo{year}{2007}\natexlab{}.
\newblock \showarticletitle{RA Fisher and his advocacy of randomization}.
\newblock \bibinfo{journal}{\emph{Journal of the History of Biology}}
  \bibinfo{volume}{40}, \bibinfo{number}{2} (\bibinfo{year}{2007}),
  \bibinfo{pages}{295--325}.
\newblock


\bibitem[\protect\citeauthoryear{Hauser, Paolacci, and Chandler}{Hauser
  et~al\mbox{.}}{2019}]%
        {HPC19}
\bibfield{author}{\bibinfo{person}{David Hauser}, \bibinfo{person}{Gabriele
  Paolacci}, {and} \bibinfo{person}{Jesse Chandler}.}
  \bibinfo{year}{2019}\natexlab{}.
\newblock \showarticletitle{Common concerns with MTurk as a participant pool:
  Evidence and solutions}.
\newblock In \bibinfo{booktitle}{\emph{Handbook of research methods in consumer
  psychology}}. \bibinfo{publisher}{Routledge}, \bibinfo{pages}{319--337}.
\newblock


\bibitem[\protect\citeauthoryear{H{\"o}st, Wohlin, and Thelin}{H{\"o}st
  et~al\mbox{.}}{2005}]%
        {host2005experimental}
\bibfield{author}{\bibinfo{person}{Martin H{\"o}st}, \bibinfo{person}{Claes
  Wohlin}, {and} \bibinfo{person}{Thomas Thelin}.}
  \bibinfo{year}{2005}\natexlab{}.
\newblock \showarticletitle{Experimental context classification: incentives and
  experience of subjects}. In \bibinfo{booktitle}{\emph{Int. conf. on Software
  Engineering}}. \bibinfo{pages}{470--478}.
\newblock


\bibitem[\protect\citeauthoryear{Juristo and Moreno}{Juristo and
  Moreno}{2013}]%
        {juristo2013basics}
\bibfield{author}{\bibinfo{person}{Natalia Juristo} {and}
  \bibinfo{person}{Ana~M Moreno}.} \bibinfo{year}{2013}\natexlab{}.
\newblock \bibinfo{booktitle}{\emph{Basics of software engineering
  experimentation}}.
\newblock \bibinfo{publisher}{Springer Science \& Business Media}.
\newblock


\bibitem[\protect\citeauthoryear{Kitchenham and Charters}{Kitchenham and
  Charters}{2007}]%
        {Kitchenham2007}
\bibfield{author}{\bibinfo{person}{B. Kitchenham} {and} \bibinfo{person}{S
  Charters}.} \bibinfo{year}{2007}\natexlab{}.
\newblock \bibinfo{title}{Guidelines for performing Systematic Literature
  Reviews in Software Engineering}.
\newblock
\newblock


\bibitem[\protect\citeauthoryear{Kitchenham}{Kitchenham}{2002}]%
        {kitchenham2002preliminary}
\bibfield{author}{\bibinfo{person}{Barbara et~al. Kitchenham}.}
  \bibinfo{year}{2002}\natexlab{}.
\newblock \showarticletitle{Preliminary guidelines for empirical research in
  software engineering}.
\newblock \bibinfo{journal}{\emph{IEEE Transactions on software engineering}}
  \bibinfo{volume}{28}, \bibinfo{number}{8} (\bibinfo{year}{2002}),
  \bibinfo{pages}{721--734}.
\newblock


\bibitem[\protect\citeauthoryear{Layman and Sigur\dh~sson}{Layman and
  Sigur\dh~sson}{2013}]%
        {Layman2013using}
\bibfield{author}{\bibinfo{person}{Lucas Layman} {and} \bibinfo{person}{Gunnar
  Sigur\dh~sson}.} \bibinfo{year}{2013}\natexlab{}.
\newblock \showarticletitle{Using Amazon's Mechanical Turk for User Studies:
  Eight Things You Need to Know}. In \bibinfo{booktitle}{\emph{Int. Symp. on
  Empirical Software Engineering and Measurement}}. \bibinfo{pages}{275--278}.
\newblock


\bibitem[\protect\citeauthoryear{Moll{\'e}ri, Petersen, and Mendes}{Moll{\'e}ri
  et~al\mbox{.}}{2016}]%
        {molleri2016survey}
\bibfield{author}{\bibinfo{person}{Jefferson~Seide Moll{\'e}ri},
  \bibinfo{person}{Kai Petersen}, {and} \bibinfo{person}{Emilia Mendes}.}
  \bibinfo{year}{2016}\natexlab{}.
\newblock \showarticletitle{Survey guidelines in software engineering: An
  annotated review}. In \bibinfo{booktitle}{\emph{Proceedings of the 10th
  ACM/IEEE Int. Symp. on empirical software engineering and measurement}}.
  \bibinfo{pages}{1--6}.
\newblock


\bibitem[\protect\citeauthoryear{Montgomery, Runger, and Hubele}{Montgomery
  et~al\mbox{.}}{2011}]%
        {montgomery2011engineering}
\bibfield{author}{\bibinfo{person}{D.C. Montgomery}, \bibinfo{person}{G.C.
  Runger}, {and} \bibinfo{person}{N.F. Hubele}.}
  \bibinfo{year}{2011}\natexlab{}.
\newblock \bibinfo{booktitle}{\emph{Engineering Statistics}}.
\newblock \bibinfo{publisher}{John Wiley}.
\newblock


\bibitem[\protect\citeauthoryear{Nagappan, Zimmermann, and Bird}{Nagappan
  et~al\mbox{.}}{2013}]%
        {nagappan2013diversity}
\bibfield{author}{\bibinfo{person}{Meiyappan Nagappan}, \bibinfo{person}{Thomas
  Zimmermann}, {and} \bibinfo{person}{Christian Bird}.}
  \bibinfo{year}{2013}\natexlab{}.
\newblock \showarticletitle{Diversity in software engineering research}. In
  \bibinfo{booktitle}{\emph{9th joint meeting on foundations of software
  engineering}}. \bibinfo{pages}{466--476}.
\newblock


\bibitem[\protect\citeauthoryear{Palan and Schitter}{Palan and
  Schitter}{2018}]%
        {PS18}
\bibfield{author}{\bibinfo{person}{Stefan Palan} {and}
  \bibinfo{person}{Christian Schitter}.} \bibinfo{year}{2018}\natexlab{}.
\newblock \showarticletitle{{Prolific.ac: A subject pool for online
  experiments}}.
\newblock \bibinfo{journal}{\emph{Journal of Behavioral and Experimental
  Finance}}  \bibinfo{volume}{17} (\bibinfo{year}{2018}),
  \bibinfo{pages}{22--27}.
\newblock


\bibitem[\protect\citeauthoryear{Robinson}{Robinson}{2019}]%
        {RRM19}
\bibfield{author}{\bibinfo{person}{Jonathan et~al. Robinson}.}
  \bibinfo{year}{2019}\natexlab{}.
\newblock \showarticletitle{{Tapped out or barely tapped? Recommendations for
  how to harness the vast and largely unused potential of the Mechanical Turk
  participant pool}}.
\newblock \bibinfo{journal}{\emph{PLOS ONE}} \bibinfo{volume}{14},
  \bibinfo{number}{12} (\bibinfo{year}{2019}).
\newblock


\bibitem[\protect\citeauthoryear{Salman, Misirli, and Juristo}{Salman
  et~al\mbox{.}}{2015}]%
        {salman2015students}
\bibfield{author}{\bibinfo{person}{Iflaah Salman}, \bibinfo{person}{Ayse~Tosun
  Misirli}, {and} \bibinfo{person}{Natalia Juristo}.}
  \bibinfo{year}{2015}\natexlab{}.
\newblock \showarticletitle{Are students representatives of professionals in
  software engineering experiments?}. In \bibinfo{booktitle}{\emph{2015
  IEEE/ACM 37th IEEE Int. conf. on Software Engineering}},
  Vol.~\bibinfo{volume}{1}. IEEE, \bibinfo{pages}{666--676}.
\newblock


\bibitem[\protect\citeauthoryear{Salvaneschi, Amann, Proksch, and
  Mezini}{Salvaneschi et~al\mbox{.}}{2014}]%
        {salvaneschi2014empirical}
\bibfield{author}{\bibinfo{person}{Guido Salvaneschi}, \bibinfo{person}{Sven
  Amann}, \bibinfo{person}{Sebastian Proksch}, {and} \bibinfo{person}{Mira
  Mezini}.} \bibinfo{year}{2014}\natexlab{}.
\newblock \showarticletitle{An empirical study on program comprehension with
  reactive programming}. In \bibinfo{booktitle}{\emph{Int. Symp. on Foundations
  of Software Engineering}}. \bibinfo{pages}{564--575}.
\newblock


\bibitem[\protect\citeauthoryear{Shadish, Cook, and Campbell}{Shadish
  et~al\mbox{.}}{2001}]%
        {Shadish2001}
\bibfield{author}{\bibinfo{person}{W. Shadish}, \bibinfo{person}{T. Cook},
  {and} \bibinfo{person}{D. Campbell}.} \bibinfo{year}{2001}\natexlab{}.
\newblock \showarticletitle{Experimental and Quasi-Experimental Designs for
  Generalized Causal Inference}. In \bibinfo{booktitle}{\emph{Houghton
  Mifflin}}.
\newblock


\bibitem[\protect\citeauthoryear{Shadish and et~al.}{Shadish and
  et~al.}{2002}]%
        {shadish2002experimental}
\bibfield{author}{\bibinfo{person}{William Shadish} {and} \bibinfo{person}{et
  al.}} \bibinfo{year}{2002}\natexlab{}.
\newblock \bibinfo{booktitle}{\emph{Experimental and quasi-experimental designs
  for generalized cusal inference}}.
\newblock \bibinfo{publisher}{WADSWORTH CENGAGE Learning}.
\newblock


\bibitem[\protect\citeauthoryear{Shepperd}{Shepperd}{2018}]%
        {shepperd2018inferencing}
\bibfield{author}{\bibinfo{person}{Martin Shepperd}.}
  \bibinfo{year}{2018}\natexlab{}.
\newblock \showarticletitle{Inferencing into the void: problems with implicit
  populations Comments onEmpirical software engineering experts on the use of
  students and professionals in experiments'}.
\newblock \bibinfo{journal}{\emph{arXiv preprint arXiv:1810.07392}}
  (\bibinfo{year}{2018}).
\newblock


\bibitem[\protect\citeauthoryear{Sjoberg}{Sjoberg}{2002}]%
        {sjoberg2002conducting}
\bibfield{author}{\bibinfo{person}{Dag et~al. Sjoberg}.}
  \bibinfo{year}{2002}\natexlab{}.
\newblock \showarticletitle{Conducting realistic experiments in software
  engineering}. In \bibinfo{booktitle}{\emph{Int. Symp. on empirical software
  engineering}}. IEEE, \bibinfo{pages}{17--26}.
\newblock


\bibitem[\protect\citeauthoryear{Sjoeberg, Hannay, Hansen, Kampenes,
  Karahasanovic, Liborg, and Rekdal}{Sjoeberg et~al\mbox{.}}{2005}]%
        {SHH05}
\bibfield{author}{\bibinfo{person}{D.I.K. Sjoeberg}, \bibinfo{person}{J.E.
  Hannay}, \bibinfo{person}{O. Hansen}, \bibinfo{person}{V.B. Kampenes},
  \bibinfo{person}{A. Karahasanovic}, \bibinfo{person}{N.-K. Liborg}, {and}
  \bibinfo{person}{A.C. Rekdal}.} \bibinfo{year}{2005}\natexlab{}.
\newblock \showarticletitle{{A survey of controlled experiments in software
  engineering}}.
\newblock \bibinfo{journal}{\emph{IEEE Transactions on Software Engineering}}
  \bibinfo{volume}{31}, \bibinfo{number}{9} (\bibinfo{year}{2005}),
  \bibinfo{pages}{733--753}.
\newblock


\bibitem[\protect\citeauthoryear{Stolee and Elbaum}{Stolee and Elbaum}{2010}]%
        {stolee2010exploring}
\bibfield{author}{\bibinfo{person}{Kathryn~T Stolee} {and}
  \bibinfo{person}{Sebastian Elbaum}.} \bibinfo{year}{2010}\natexlab{}.
\newblock \showarticletitle{Exploring the use of crowdsourcing to support
  empirical studies in software engineering}. In \bibinfo{booktitle}{\emph{Int.
  Symp. on Empirical software engineering and measurement}}.
  \bibinfo{pages}{1--4}.
\newblock


\bibitem[\protect\citeauthoryear{Stolee and Elbaum}{Stolee and Elbaum}{2013}]%
        {stolee2013use}
\bibfield{author}{\bibinfo{person}{Kathryn~T Stolee} {and}
  \bibinfo{person}{Sebastian Elbaum}.} \bibinfo{year}{2013}\natexlab{}.
\newblock \showarticletitle{On the use of input/output queries for code
  search}. In \bibinfo{booktitle}{\emph{Int. Symp. on Empirical Sof. Eng. and
  Measurement}}. \bibinfo{pages}{251--254}.
\newblock


\bibitem[\protect\citeauthoryear{Stolee, Elbaum, and Dobos}{Stolee
  et~al\mbox{.}}{2014}]%
        {stolee2014solving}
\bibfield{author}{\bibinfo{person}{Kathryn~T Stolee},
  \bibinfo{person}{Sebastian Elbaum}, {and} \bibinfo{person}{Daniel Dobos}.}
  \bibinfo{year}{2014}\natexlab{}.
\newblock \showarticletitle{Solving the search for source code}.
\newblock \bibinfo{journal}{\emph{Trans. on Software Engineering and
  Methodology}} \bibinfo{volume}{23}, \bibinfo{number}{3}
  (\bibinfo{year}{2014}).
\newblock


\bibitem[\protect\citeauthoryear{Stolee, Elbaum, and Dwyer}{Stolee
  et~al\mbox{.}}{2016}]%
        {stolee2016code}
\bibfield{author}{\bibinfo{person}{Kathryn~T Stolee},
  \bibinfo{person}{Sebastian Elbaum}, {and} \bibinfo{person}{Matthew~B Dwyer}.}
  \bibinfo{year}{2016}\natexlab{}.
\newblock \showarticletitle{Code search with input/output queries:
  Generalizing, ranking, and assessment}.
\newblock \bibinfo{journal}{\emph{Journal of Systems and Software}}
  \bibinfo{volume}{116} (\bibinfo{year}{2016}), \bibinfo{pages}{35--48}.
\newblock


\bibitem[\protect\citeauthoryear{Vanhanen and Lassenius}{Vanhanen and
  Lassenius}{2005}]%
        {vanhanen2005effects}
\bibfield{author}{\bibinfo{person}{Jari Vanhanen} {and} \bibinfo{person}{Casper
  Lassenius}.} \bibinfo{year}{2005}\natexlab{}.
\newblock \showarticletitle{Effects of pair programming at the development team
  level: an experiment}. In \bibinfo{booktitle}{\emph{2005 Int. Symp. on
  Empirical Software Engineering, 2005.}} IEEE, \bibinfo{pages}{10--pp}.
\newblock


\bibitem[\protect\citeauthoryear{Vegas, Dieste, and Juristo}{Vegas
  et~al\mbox{.}}{2015}]%
        {vegas2015difficulties}
\bibfield{author}{\bibinfo{person}{Sira Vegas}, \bibinfo{person}{Oscar Dieste},
  {and} \bibinfo{person}{Natalia Juristo}.} \bibinfo{year}{2015}\natexlab{}.
\newblock \showarticletitle{Difficulties in running experiments in the software
  industry: Experiences from the trenches}. In \bibinfo{booktitle}{\emph{Int.
  Workshop on Conducting Empirical Studies in Industry}}. IEEE,
  \bibinfo{pages}{3--9}.
\newblock


\bibitem[\protect\citeauthoryear{Wohlin and et~al.}{Wohlin and et~al.}{2012}]%
        {wohlin2012experimentation}
\bibfield{author}{\bibinfo{person}{Claes Wohlin} {and} \bibinfo{person}{et
  al.}} \bibinfo{year}{2012}\natexlab{}.
\newblock \bibinfo{booktitle}{\emph{Experimentation in software engineering}}.
\newblock \bibinfo{publisher}{Springer Science \& Business Media}.
\newblock


\end{thebibliography}



\end{document}